# Scientific progress despite irreproducibility: A seeming paradox

Richard M. Shiffrin[a,1], Katy Börner[b], Stephen M. Stigler[c]

[a] Department of Psychological and Brain Sciences, Indiana University, Bloomington, IN 47405

[b] Department of Intelligent Systems Engineering, Indiana University, Bloomington, IN 47405

[c] Department of Statistics, University of Chicago, Chicago, IL 60637 USA

[1] Corresponding Author: Email: shiffrin@indiana.edu

**Abstract**

It appears paradoxical that science is producing outstanding new results and theories at a rapid rate at the same time that researchers are identifying serious problems in the practice of science that cause many reports to be irreproducible and invalid. Certainly the practice of science needs to be improved and scientists are now pursuing this goal. However, in this perspective we argue that this seeming paradox is not new, has always been part of the way science works, and likely will remain so. We first introduce the paradox. We then review a wide range of challenges that appear to make scientific success difficult. Next, we describe the factors that make science work—in the past, present, and presumably also in the future. We then suggest that remedies for the present practice of science need to be applied selectively so as not to slow progress, and illustrate with a few examples. We conclude with arguments that communication of science needs to emphasize not just problems but the enormous successes and benefits that science has brought and is now bringing to all elements of modern society.

**Significance Statement**

The 'crisis' of reproducibility gives a view of present day science that is highly negative and may give ammunition to those who would want to deny the findings of science and reduce funding. Science is in fact proceeding remarkably well and remarkably quickly. Our article uses this 'paradox' framing, and some historical perspectives, to explain that irreproducibility and invalidity of some reports are a natural part of scientific practice, and to show how rapid progress occurs despite the problems. We hope the submission will provide a more balanced view of present day science, and provide a perspective that will help communicate science and its benefits to the public.

# 1. A Seeming Scientific Paradox: Progress and Problems

What appears to be a paradox is evident when evaluating science today: It is producing at a dizzying rate findings that are of fundamental importance for theory and of extraordinary benefit for society. Yet scientists are engaged in an extensive effort of self-evaluation, identifying and addressing problems in the practice of science that can produce reports that are invalid or cannot be reproduced.

This article, a mixture of a perspective, an opinion piece and an editorial, aims to shed light on this apparent paradox of success in the face of serious challenges.  It places the present practices of science in an historical perspective, highlighting how science has worked historically and is working today.

It would require many books to lay out the vast number of recent and important scientific advances in theory, understanding, and applications. In fact, there are so many advances in so many diverse fields occurring so rapidly that it is difficult to find a source that manages to track them all, even within a given field. A variety of magazines and newspapers cover a tiny fraction of the more interesting advances for the lay public. Key examples are captured in *Time 100 New Scientific Discoveries: Fascinating, Momentous, and Mind Expanding Breakthroughs* [1], *New York Times*' Tuesday "Science Times," *Wall Street Journal*'s roundup of the year's top science stories, *Christian Science Monitor's* listing of outstanding scientific breakthroughs, or *WIRED's* highlighting of major breakthroughs. A slightly larger and more detailed list of advances is found in reports from the National Academies of Sciences, Engineering, and Medicine, the National Science Foundation (NSF), the National Institutes of Health (NIH), and the American Association for the Advancement of Science (AAAS); these publish a selection of major scientific, health, and engineering advances in some detail for what is likely a target audience of scientists. No one aware of the present state of science doubts the many important advances that are taking place almost daily.

Although there is good reason to believe in the rapidity of scientific progress, we know of no feasible methods to produce a quantitative metric, either across science or within field. Proving the claims of rapid progress would be inordinately difficult and beyond the scope of this contribution.  If say 10% of this year's publications were valid, important and useful, such a proof might require a first analysis of some 2,000,000 publications (in diverse fields) to obtain a candidate set of 200,000, followed by detailed analysis of those selected. Trying to reduce this scope by sampling would require general agreement about the structure of science (e.g. the greatest progress is likely induced by a relatively small proportion of publications). Even if it were possible to carry out such a procedure, it would still be inadequate because it often takes many years for the validity, importance and utility of contributions to become apparent.  In addition, whatever metric could be imagined it would surely differ by discipline. Partly for these reasons, scientists and the public should not be led to assume that progress is limited to some fields and not others. The reports of reproducibility tend to focus on certain fields (e.g. some social sciences because the reports resonate with readers' intuitions, and some biological sciences because of their importance for health), but such choices should not be taken to imply lack of progress, or even differential progress in comparison with other fields.

Given that advances in science are generally evidenced by report of a few salient cases, and given the difficulty of measuring progress generally, is there any way one might get a sense of the scope of the advances? If even a small proportion of reports are indeed valid and contribute to scientific progress, the numbers are very high. Consider just the reports given at the annual meeting of one scientific subfield, the Society for Neuroscience: If just 10% of the roughly 6,000 research presentations were valid and important, that would equate to between one and two valid and important results per day in just this one subfield. The numbers are similar across the scientific spectrum. Of course the 10% estimate is not verifiable, but almost all scientists would think this to be an underestimate, and even if a slight overestimate, the numbers of good studies would be immense. The public—and to a large extent scientists themselves (except in their narrow domains of expertise)—are not usually aware of scientific advances but rather the translation of the evolving web of scientific advances into applications that increasingly help them in their daily lives. Examples are particularly evident in technology and health:  computers, cellphones, satellites, GPS, information availability, antibiotics, heart operations, increasing age of mortality, fertilizers, genetically improved crops, minimally invasive joint surgeries and joint replacements, contact lenses and laser surgery, magnetic resonance imaging, robotics, electric vehicles, air bags, solar energy, wind energy, better weather forecasting, and on and on almost without end.

At the same time, no one can deny the evidence, collected by scientists themselves, that there exist serious problems in the practice of science. Scientists know that many of these problems can be remedied, or at least improved, and are working to that end. Key issues were highlighted in the presentations at the Sackler Colloquium "Reproducibility of Research: Issues and Proposed

Remedies", and are discussed in other articles in this special issue [2]. It is not the aim of this perspective to justify poor practices—we are in complete agreement with our colleagues that such practices need to be remedied, keeping in mind the need to insure that the remedy does not do more harm than the disease. Nonetheless, rapid scientific progress is occurring despite the problems and we endeavor in this article to resolve the seeming paradox by reviewing the way that the practice of science has evolved and is now working. Subsequent sections identify key challenges to scientific progress, discuss the self-correcting inner workings of science, and debate the pros and cons of some proposed remedies. We conclude with a discussion of the need to communicate more clearly scientific progress and science's benefits for society.

## 2. Challenges to Scientific Progress

While irreproducible and invalid results are the main focus of this paper, and are discussed next, we also describe other challenges that present major obstacles to scientific progress. All of these challenges make the successes of science seem paradoxical.

### 2.1 Challenges of Reproducibility and Validity

There exist serious problems in the practice of science that lead to the publication of irreproducible and invalid results. One problem is an incentive structure that rewards quantity of publications over quality, hasty rather than thoughtful publication, report of unlikely results, and research biased toward current areas of funding. Other problems include: withholding of relevant data; selective reporting of positive rather than negative results; toleration of too weak statistical criteria for publication; error prone statistical analyses; inadequate reporting of design, analysis, data, and computer programs and codes; unconscious biases that distort empirical design, data analysis and reports; post-hoc selection of findings to report; inadequate reviewing.

Whether or not the problems of reproducibility should be described as a crisis (the record of scientific progress may suggest other terminology) we note that the current focus on reproducibility misses what is likely an even more important concern: the validity of reports. First, we know that reproducibility does not guarantee validity: All scientists know many examples of invalid results and conclusions that are reproduced many times, mostly due to repetitions of the problems mentioned in the previous paragraph (ESP studies provide just one example). Second, invalid reports can do more harm than irreproducible reports: A weakly supported claim that societal structure imposes an implicit bias against certain minorities does less harm than a weakly supported claim that it does not. Finally, we note that validity is desirable but even more is needed to maximize progress: Validity refers to a qualitative aspect of reporting, for example reporting the true 'direction' of a finding when one reports a simple effect justified by null hypothesis testing. However, an effect valid in only one exact context and not others is not usually useful and important. We want to report effects that are reliable, important, of scientific value, and that generalize to similar settings. Problems of scientific practice do not just produce irreproducibility; they impair all these goals.

## 2.2 Information Overload

Science, along with all other elements of society, is now in the era of Big Data. The amount of data being collected, the number of scientific journals, science publications, science blogs, and scientific information generally, has outstripped human ability to comprehend and act upon those data. We have sophisticated computational tools to sift information and point to publications likely to be relevant, but at the end of the pipeline sits an individual scientist with limited time. Any scientist who has taken days to review a single submission is aware of the scope of this problem. Faced with this ever growing problem, scientists have become good at using their scientific and statistical judgment to filter publications on the basis of limited information, such as abstracts and information gleaned from rapid scanning (a skill that helps resolve the paradox, as we discuss later). It is not new that scientists feel information is growing faster than they can monitor effectively: It is likely that scientists in every generation have been aware that the breadth of knowledge they were able to encompass when in training took in a larger proportion of their field than was the case when their careers had matured. Nonetheless, the rapid growth we have seen in the information age, as technology allows us to collect and store previously unimaginable amounts of data, magnifies this problem and brings it to the forefront. We illustrate some aspects of this problem in Figure 1, showing the vast increase in scientific publications that has taken place from 1800 to the present [3]. Most of this exponential increase has occurred in the last few years and appears to continue. How can science progress when the critically important findings seem to be lost in a sea of scientific reports, some valid and some invalid?

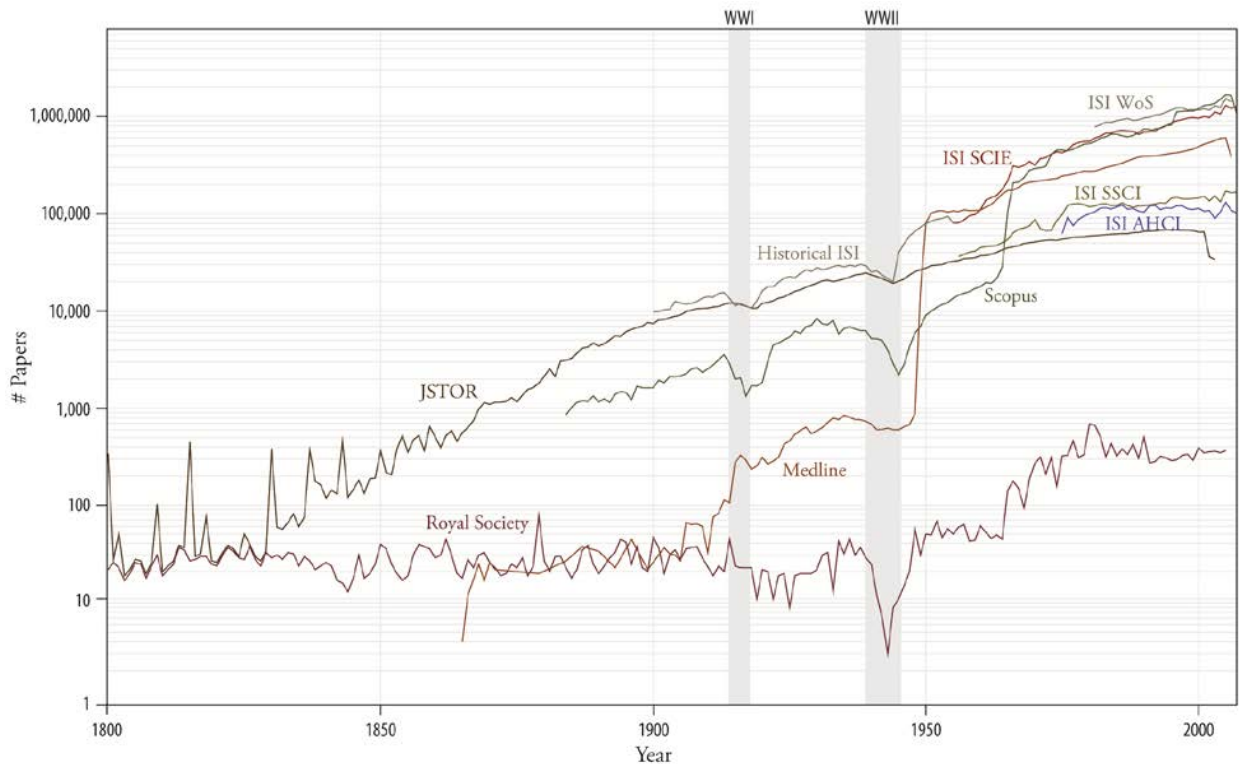

**Figure 1. Accelerating Growth of Publications.** The graph shows the number of papers (log scale) per publication year from 1800 to 2007 based on numerous data sources, see labels. The growth in papers is dominated by increases in recent years. Note the impact of World Wars I and II on scholarly productivity. Adopted from [3].

### 2.3 Increasing Specialization

Closely related to the growth of knowledge is the increasing specialization of that knowledge, and the technical expertise needed to gain mastery of even a narrow domain. It may seem surprising that progress continues unabated when a large part of a scientist's professional life is spent in training to obtain the expertise needed to make contributions. To the extent that younger scientists are the engine of creativity, increasing specialization may reduce progress by delaying the age at which they can contribute. In addition, increasing specialization is very likely a causal factor in the increasing age at which investigators can obtain funding. The average age at which researchers receive their first NIH R01 grant and tenure-track jobs have both increased, e.g., in 1980, close to 18% of all PIs were age 36 and under; that number had fallen to about 3% in 2010 [4]. Progress may also be hindered when younger scientists cannot obtain funding early in their careers [5]. To counter the problems caused by increasing specialization scientists of all ages have increasingly engaged in collaborations, as discussed next.

## 2.4 Collaborative Science

Science is increasingly performed via large-scale collaborations, partly in response to increases in the specialization of knowledge. This is evidenced by the decrease in the number of single authored publications and the increase in average number of authors [6]. It is also the case that the collaborations are increasing in distance and across countries. For example, Figure 2 shows the dense network of international co-affiliations for faculty at a single university, Indiana University, over a recent ten-year period using publication data from Web of Science, Clarivate Analytics [7, 8]. The scope of collaborations is hinted at by the figure—the full co-author network would be much denser and impossible to graph in the space available here. International, long-distance, and long-term collaborations of course introduce their own problems, particularly of reproducibility, validity, and costs.

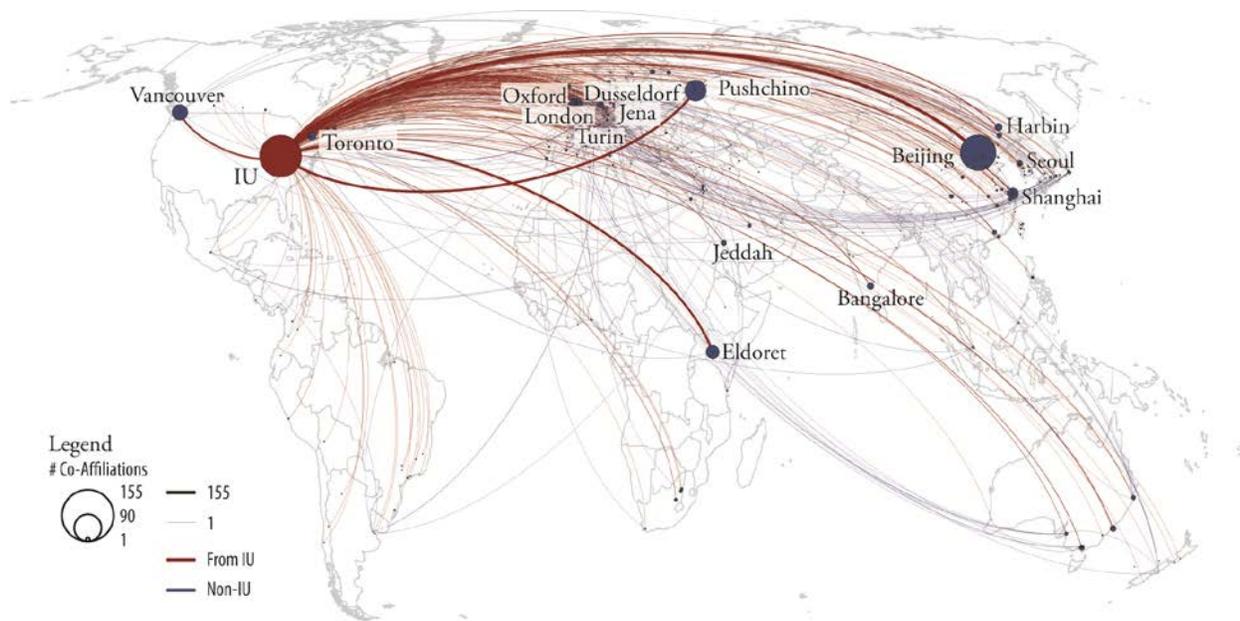

**Figure 2. Global Co-Affiliation Network.** This world map shows the co-affiliations of authors that listed "Indiana Univ" and at least one other non-U.S. institution as affiliation on 1,590 scholarly papers. Publication data was retrieved from the IUNI Web of Science database and covers 2004-2013. There are 344 affiliation locations (not counting IU) and 641 co-affiliation links. Nodes denote author locations and are area size coded by degree (i.e., the number of links per node) with the exception of IU which has 1,592 co-affiliation links. Links denote co-affiliations, e.g., an author with three affiliations IU, X, Y will add three links; the two links that connect IU with X and Y are drawn in red while the link between X and Y is given in purple. Links are thickness coded by the number of co-affiliations.

Today's science has grown in many ways: more scientists, more data, more complex research problems, more collaboration. With this change in scale, problems that in the past could be dealt with easily become less tractable. The growth in collaboration may be necessary to accumulate needed expertise, but brings with it a division of labor and a potential loss of individual responsibility. The growth in complexity and collaboration makes exploration multifaceted and assessment and control difficult.

## 2.5 Research Cost Inflation

Yet another factor operating against scientific progress is the rapidly increasing cost of many domains of research, an increase that is not generally matched by increases in funding. Figure 3 gives an example of how costs have increased, showing the average costs for all publications in a given year for research on penicillin (red) and CRISPR (black). The substantial costs involved reflect the personnel, infrastructure, materials and equipment, data and software development efforts required to make progress. The costs of course vary by field but are rising in all fields of science. Some important research efforts require truly enormous amounts of support from many entities. Examples include large-scale infrastructures such as the Large Hadron Collider (LHC) or the Giant Magellan Telescope (GMT). The costs to develop new drugs to treat disease are another well-known example. Again it seems paradoxical that rapid progress has been occurring when the costs of research threaten to outstrip available funding.

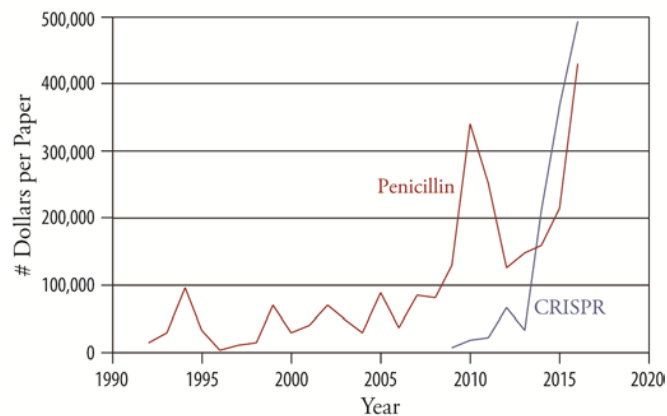

**Figure 3. Increase in Research Costs Per Published Paper.** The graph shows the increasing costs of research for two biological research areas, penicillin and CRISPR. Data on NIH awards and publications that acknowledge these grant awards was retrieved from the Research Portfolio Online Reporting Tools (RePORT) [9]. Research costs for all publications are plotted for each publication year. As can be seen, costs have gone up substantially over the recent years; costs for CRISPR have increased dramatically. Note that there is a delay between funding awarded in 2015 and publications being published that acknowledge this funding, e.g., the 2015/2016 values will likely decrease as future publications acknowledge this past funding.

### 2.6 Administrative Overhead

Researchers today face increasing numbers of rules and regulations at every level, federal and state, university and business in preparing, submitting, and reporting grants, respecting privacy and health regulations, obeying well-intentioned laws and rules that seem always to increase and add to the administrative burden. Ironically, some of the remedies proposed to deal with the reproducibility crisis will inevitably increase the research overhead, although one would hope the benefits to science would outweigh the costs.

## 3. The Practice of Science

The challenges discussed in the previous section highlight the difficulty of achieving significant scientific progress. Yet, scientific progress is occurring, and rapidly. In this section, we start by noting that most of these issues have always been at the core of science, albeit in different degrees. Scientific practice has therefore evolved to deal with the problems. We then discuss some of the key factors that have helped and now help science to succeed despite poor practices, information overload, increasing specialization, underfunded research, and the other factors that make research progress challenging.

### 3.1 Learning from The Past

Science works in complex ways. The issues concerning reproducibility are not new to science. Examining how science has worked in the past, has evolved, and is working today helps to understand the paradox. Historians of science naturally focus on the great successes or the spectacular failures, not the great majority of reports that in many cases form the necessary substrate upon which the spectacular advances build, and in other cases are invalid, irreproducible, and/or unimportant, and are largely ignored. Yet these large numbers of historically ignored research results have been published in the scientific literature since publications first appeared and are—we will argue—a necessary component of exploration. We are not aware of any study that aims to assess whether the many low impact and ignored publications in earlier epochs of science were valid or reproducible, and to assess whether the proportion of such reports is now higher than was true historically. However, there are sufficient numbers of well-studied cases involving invalid findings to bolster the claim that such reports have been a normal component of scientific investigation starting at the time that the scientific method became the dominant approach. A few illustrative examples are given here.

In some historical cases, a false result seemed at first to be reproducible, yet failed a later and more careful investigation. The Greeks believed it rained after major battles, as the gods washed the blood of the fallen off the field of battle. The belief took a modern cast in the 19th century when it was believed the noise of the battle, e.g., made by cannons, would cause the clouds to release rain; by the 1880s, it was thought this could be a way to answer farmers' prayers. A survey of officers serving in the Civil War led to a book length set of replications, as one after another when asked if they could recall a battle where it rained soon after gave supporting evidence. Congress allotted money for a study in Texas; cannons were fired and a lack of reproducibility closed the case. The false theory was supported by many of the problems highlighted by recent investigators of irreproducibility: The initial investigators failed to ask critical questions and failed to report critical data: How soon after a battle? How much counts as rain? Over what area? Reports were selective: no one was asked for battles with no rain after [10]. One or two other examples are well known: Even into the 20th century, astrophysicists believed "empty space" was filled with a "luminiferous aether" that carried light over vast distances. Some noted astronomers refused to abandon the notion of canals on Mars even in the early 20<sup>th</sup> century, well after the source of the claimed observation—imprecision in telescopic lenses—had been noted [11].

On the other hand, there are many cases where a true theory at first produced no confirmation. In the 1820s, Laplace knew that, in theory, there should be tides of the atmosphere just as there are tides of the sea, and he set to measure the atmospheric tide using several decades of data on barometric pressure at the Paris Observatory taken each day at 3 hour intervals. Despite an unusually large amount of data and a clever design for testing (using within-day variation that blocked out day-to-day variation), he still failed to detect the tide. It later turned out that Paris was a terrible site to choose: the tide there (while real) was barely perceptible. A different location could have given the desired result [12].

Karl Pearson discovered the fetal alcohol syndrome in 1911 but he then ignored it because the effect was smaller than the effects he sought and not what he was looking for. Later, researchers without Pearson's narrow focus saw it differently and made the discovery that Pearson missed [13]. As Alfred North Whitehead wrote [14], "To come very near to a true theory, and to grasp its precise application, are two very different things. Everything of importance has been said before by somebody who did not discover it."

### 3.2 Science Requires Exploration

To shed light on the paradox of scientific productivity despite irreproducibility and other challenges that hinder research, it is useful to distinguish exploratory science from translational science. Exploratory science can be defined as empirical and theoretical discoveries. Translational science is commonly defined as the process that turns discoveries into applications intended to benefit society or businesses. There is of course no hard boundary between these approaches to research. In reality, there are so many forms of research, varying on almost uncountable dimensions, that any binary classification is at best a crude and nominal approximation. Occasionally, the presence of translational research is clear, as when a researcher carries out a field test of a potentially useful drug. In other cases, the research may have been exploratory in character, say from a laboratory study, but then is used as a basis for an extended series of studies used to produce an application. Many cases lie somewhere along this continuum. However, costs for these two extremes differ: in translational research, the costs of mistakes can be very high, as in health treatments or business decisions that may be critical to success. In exploratory research the costs tend to be lower, sometimes limited to resources temporarily expended pursuing false leads.

We turn next to exploratory research where publications of promising but possibly invalid results are a natural basis for scientific progress. Exploratory science is a matter of knowledge accretion, often over long periods of time. A metaphor would be a constantly growing tree, occasionally sprouting new and large branches, but one that continually modifies many of its previous branches and leaves, as new knowledge develops. Invalid and irreproducible reports are buds on the tree that have the potential to sprout but occur in positions that do not allow growth. The 'tree' does not know which positions foster growth, and therefore forms many buds, some failing and others sprouting.

This metaphor suggests why invalid and irreproducible reports appear on the scientific tree. Science is at heart a matter of sophisticated trial and error. In most exploratory science the assessment of results will depend crucially upon detailed models of experimental variation that cannot be defined in advance, so promising results are pursued in further research that eventually determines validity, reproducibility, importance, and utility. Many avenues are explored to increase the probability of including the few of critical importance.

To carry these thoughts further, we note that scientific advances very often do not appear in full and final form when first published but rather require a period of maturation fostered by additional research. Thus it is advantageous for science to include publication of many promising results, even though some prove invalid, because the validity and importance of results often does not become clear until much time has passed. Such findings are sometimes termed 'sleeping beauties' [15], and highlighted by the *Golden Goose Awards* sponsored by a wide variety of organizations interested in scientific progress. Sleeping beauties, and valid and useful reports generally, are of course mixed among many other reports that do not prove useful, but it takes time and additional research to know which are which. To say this another way, it is useful to publish promising results, though some prove invalid, because scientific discovery is necessarily rooted in unpredictability.[*]

### 3.3 Balancing Creativity and Rigor

To maximize scientific progress, one must find the right balance between free rein to publish promising results at an early stage and overly strict criteria for publication. Even to speak of "criteria" is a fairly modern notion. For centuries these criteria were not articulated but were buried in a scientific community's general acceptance, applied differently in different communities and at different times. William Harvey's wonderful experiment conclusively demonstrated the circulation of blood in 1628, but he was initially greeted by two opposed reactions: "It isn't true" and "It has been known since Hippocrates"—history does not seem to record any contemporary attempts to replicate his work. Only after more discussion and consideration was it generally accepted, and even then not by any explicit criteria [17]. In general, an appropriate balance of creativity and rigor will mandate occasional publications of invalid and irreproducible results, and such a balance will be needed to optimize the rate of scientific progress.

### 3.4 Scientists are Skeptical

Scientists are keenly aware, and have always been aware that science is a noisy process; that reports are not always reliable; that initial conclusions might be later overturned or replaced by better ones; that science proceeds in trial and error fashion; that science constantly moves and grows—gradually improving our understanding of what is likely an infinitely complex universe. Alfred Marshall, who was said to be responsible for bringing mathematics into economics, wrote in 1885: "The most reckless and treacherous of all theorists is he who professes to let facts and figures speak for themselves, who keeps in the background the part he has played, perhaps unconsciously, in selecting and grouping them." [18, p. 202].

In 1986, when the discovery of high-temperature super-conductivity was announced there was widespread surprise and skepticism. When several others replicated the finding (which is still not well-understood) the skepticism melted away. When a respected scientist claimed in 1989 to have produced signs of cold fusion in his lab, a possibility discounted by most of the profession, there were many disbelieving responses, but the profession as a whole was respectful (perhaps recalling super-conductivity), and conferences and other labs investigated, all without any confirmation of the result. The claim was eventually dismissed as a result of lab contamination. These are just two examples that serve to illustrate how scientists have adapted their practice of science to operate effectively in the face of uncertainty [19, 20]. [†] To repeat: Scientists are at heart skeptical, and do not accept reports at face value, but rather as suggestive guides to further research and eventual applications.

### *3.5 The Distribution of the Value of Contributions*

Which studies are critically examined can lead to different conclusions about the state of science: If a "bad study" is the focus, it will highlight problems and a "good study" will highlight progress. One might think it appropriate to assess the current state of science by selecting 'randomly' a set of studies. However, such selection will produce a distorted assessment because a distribution of scientific contributions by type is almost certainly asymmetrical: It is our belief that a snapshot at a given moment in time will show that a small proportion of studies is producing the majority of scientific successes (albeit large in absolute numbers). The great majority of studies will at a given moment lie in an uncertain state of importance and validity, awaiting further research and development. In the absence of a good or feasible metric for assessing quality of publications, it would not be possible to determine these proportions with any precision. Scientific studies range from a mixture of unimportant, promising, valid and invalid, to clearly worthwhile, to important and noteworthy, to breakthroughs, with the proportions likely decreasing across these categories. Which studies are which is not necessarily known at the time of initial publication. The limitation on scientific resources (e.g., costs, time, and personnel) will prevent research follow-up of all the studies that begin in an uncertain state of importance and validity. Scientists therefore assess such studies based on their expert judgment honed by years of training; they ignore some studies but incorporate, replicate or build on others. If one imagines tracking studies over time, those studies that prove valid and valuable move on to becoming important and noteworthy, and a few become true breakthroughs. The speed of movement of results and theories can be very rapid or quite slow (as in the case of 'sleeping beauties'). All studies that prove valid and useful form the fabric of scientific knowledge, and provide the basis for further progress. Those studies that are rejected upon first inspection, and those that don't prove valid or useful are mostly ignored thereafter and forgotten. Supporting evidence for some of these claims comes from the highly skewed distribution of citation counts [21]; about half of all papers are never cited while relatively few have 100,000 citations or more; in 2014, the most highly cited paper had 305,000 citations [22]. While it seems desirable to reduce the proportion of invalid studies, a large set of promising studies, some of which will prove invalid, serve to 'pump prime' scientific progress. [‡]

This view goes a long way to unfolding the seeming paradox we have been discussing: A focus on the great mass of studies that have just been published and lie in an uncertain state of importance and validity might lead to the conclusion that science is failing, whereas a focus on those studies that are recognized and are validated as outstanding would lead to a conclusion of remarkable progress. We repeat that the latter group might seem low in proportion, but with more than two million scientific studies published each year, the absolute number of critically important results and scientific breakthroughs will surely be considerable, and sufficient to produce the rapid progress we see in science and sufficient to produce the benefits that we utilize daily in modern society.

### 3.6 The Scientific Social Network

With more than two million scientific publications each year, encompassing very large numbers in almost all disciplines, it is not possible for scientists to read all or most of the relevant published literature. Even subfields publish more papers than one human can read; e.g., neuroscience published 340,210 papers in the years 2006-2015 [24]. Web based search algorithms can help locate relevant papers, but such algorithms cannot judge quality of research, and the number located by search algorithms can be too large to allow careful reading of each. At one time the careful review at a small number of top journals gave the imprimatur of quality to a small number of papers, but with the huge growth in literature that guarantee has been weakened. What are needed are pointers to quality research. Such pointers occur in the social network of scientists, not just in the form of web-based interchange, but in the form of meetings, workshops, colloquia, schools and summer schools, and in the rich in-person communication that occurs in such venues.

### 3.7 Proceeding by Trial and Error

Studies that seem promising are replicated, built upon, or extended by others. Many are rejected during this process. The studies that can be replicated are the ones that are valued and form the basis for progress. They are the studies that make us "stand on the shoulders of giants"—former generations of scientists that laid the foundation for today's scientific progress.

In exploratory science, lack of validity does not stop learning. Scientists do not accept reports at face value. When they hear a result that looks interesting, important, and promising, they apply a series of internal filters: first to decide whether the report is likely valid and worth further exploration; second, does the work appear valid and sufficiently important to invest in further research. Most results do not pass this second filtering. Most irreproducible studies are filtered out by these two stages. The valid studies that survive are the ones that lead to steady progress. These filtering stages are a key part of the error-correcting nature of scientific practice. The many reports that initiate this filtering process are a necessary part of science.

### 3.8 Science by Combat

En route to publication, scientists work hard to correct their own findings and conclusions, knowing the high costs of 'losing reputation' when presenting false results. However, once having reached the point of publication, the extensive investment of their own human capital can make scientists defensive. They tend to resist charges of error. Thus, scientists may fail to contribute to the error correction process of science when it involves their own research. [§]

However, scientists are highly competitive, and promising results are almost always picked up and questioned by other scientists who carry out their own research, often pointing to other results and conclusions by identifying problems in the initial studies. This can be termed 'science by combat' and has been a hallmark of scientific exploration since the dawn of the scientific method. Newton deduced that the shape of the Earth should be an oblate spheroid (flattened at the poles) from his gravitational theory. The French astronomer Cassini took a contrary view on a less well-founded principle, and expeditions were mounted to Lapland and Peru to make measurements that could settle the issue. Newton was crowned the winner, albeit posthumously, illustrating the mutability of scientific advance, and the time that is often required for science to 'settle' an issue under debate [12]. Invalid and irreproducible studies tend to be weeded out as such combat proceeds, even though considerable time may be required for this process to converge. The valid ones tend to win and contribute to the progress we see.

### 3.9 Optimizing Return on Investment

Given limited resources and the need for much exploration before findings can be translated into applications, researchers commonly start with relatively low-expense, low-power studies that point in promising directions. They reserve more expensive, more careful, and more systematic exploration for subsequent research for which the extra expense can be justified by the earlier results. The initial studies are less likely to be reproducible and valid, but are useful to justify the more expensive, possibly translational, studies that follow. This way of doing science is analogous to venture capitalists investing in innovation funding on the basis of some promising application that might or might not pan out.

## 4. Targeting Remedies

Our discussion of how science works to produce progress in the face of serious challenges suggests that we adopt a nuanced approach when changing the practice of science, not one that assumes 'one size fits all'. Care is needed lest reforms and remedies have unintended consequences that unnecessarily slow the current high rate of progress. We illustrate the need for nuance with a few examples where unintended consequences might hide.

***Standards for Exploration and Exploitation***: It seems sensible to adopt "precision remedies" tailored to the type and goals of research. It seems right to use somewhat looser remedies and standards for purely exploratory research, so as to allow creativity and innovation to flourish. The case for different standards is bolstered by consideration of the differing costs of errors. The costs of publishing an invalid result during scientific exploration likely causes little harm. A faulty application could bring serious damage to health or business.

***Publication Criteria***: Different fields of science might well find it important to use different criteria for publication. At present, different fields use different standards, due to differing precision of measurement, and differing control of variables that affect results. In some fields, scientists argue persuasively that present criteria are too lax and should be strengthened. For example, there are fields that have used as a criterion that a single test achieve a 5% level of significance, defined by traditional null hypothesis testing.[#] It is now generally agreed, even by those who would like to retain null hypothesis testing, that this criterion can, with poorly planned or executed studies, promote an unacceptable level of reproducibility. Of course, statistics itself is evolving and is presently producing new and alternative methods, such as Bayesian approaches, with different criteria. There is danger in any attempt to apply a 'one size fits all' set of rules.

***Establishing Replications in Scientific Practice, and Publishing Failures***: It is at least implicit in much of the recent discussion of reproducibility that replication studies ought to be a point of stronger emphasis in the practice of science. Historically, replications have been relatively rare. Most follow up research has always and is now aimed to extend and generalize rather than replicate, confirmation or rejection falling out in the process. If additional resources are expended to carry out increased numbers of replication studies, and failures are published, validation might well be increased, but the diversion of resources could slow progress. In addition, exact replications are difficult or impossible to execute. A well-designed failure to replicate might indicate a failure to generalize to a very similar setting, but not invalidity of the original report. Generalization is often a key to scientific progress. Thus the rate of progress might well be facilitated more by studies adding conditions to explore generalization than by simple demonstrations that a report does not replicate.

A number of scientists have proposed mechanisms to foster editorial practices that will less often reject submissions of non-replications. We agree: The natural tendency of scientists to protect their own prior claims, a natural component of science by combat, may too often prevent publication of new research aimed to counter earlier claims, at least in the most respected journals. To what degree this tendency harms progress is unclear. Even when research akin to replication is undertaken, scientists often do not find the failure to confirm worth reporting. Such decisions occur for a variety of possibly good reasons: The researcher can conclude that the issue is unusually complex with no simple conclusions, that the issue is too dependent on the precise initial conditions, that establishing an alternative view would require too many scarce resources, that the results or conclusions turn out to be far less important than originally argued, that there are too many deviations from the original study, or that the new research is not executed well enough.

***Preregistration of Studies*:** There are a number of proposals that scientists should pre-register studies as a condition for publication (e.g., Nosek, et al., this issue [27]). Among other justifications is the desire to reduce post-hoc analyses and cherry-picking data and conclusions drawn from them (see below). There are many variants of how to best accomplish the pre-registration process and the extent to which this would be valuable might well differ for translational and exploratory research. Scientists are not omniscient and carry out exploratory research because they do not know in advance what they might find, nor even what kind of design and analysis might best reveal new important avenues. Whatever form of pre-registration is suggested, it is important that it not reduce creativity in exploration. The present system of science, with all its faults, uses exploration to great effect.

***Post-hoc Selection of Findings*:** Related to the issue of pre-registration are arguments against report of findings noticed only after examination of rich data sets. Scientific advances are often driven by unexpected findings, often praised as 'serendipity'. Statistical issues aside, serendipity has always been and always will be a major driver of scientific progress. In an ideal world the scientist might follow up serendipitous findings before publication, but a variety of factors, such as costs, might make this difficult or impossible.

***Survival of the Fittest:*** We have tried to show that rapid progress operating at the same time as there exist serious problems in the practice of science is not a paradox, but rather a natural outcome of the way that science has always worked, is presently working, and likely always will work. This raises concerns about the extent of reforms that one might wish to impose. Is it possible that we have to tolerate certain problematic practices (such as publication of underpowered but promising research) in order to optimize progress? We believe history demonstrates that scientific ideas flourish when they are the result of a "survival of the fittest." Such an evolutionary view requires that there exist many (also unfit) attempts that are tried and then discarded. In fact, it is probably inevitable that such an approach be used: Both current funding levels and limited resources of time, equipment, and personnel insure that exploratory science cannot be conducted at a level of sophistication that ensures complete reproducibility. Thus many low-cost and less reproducible studies are the first stage of scientific exploration, the approaches that fail are weeded out, and the valuable advances survive and are pursued. This way of viewing matters makes the paradox dissolve: Science proceeds by trial and error, and one must accept the inevitability of error while doing what is possible to minimize it.

This idea is reinforced by drawing a partial analogy with evolutionary processes: Every member of every living species is descended from a long unbroken line of survivors. We depend upon that, not only for our survival, but also for our vitality and adaptability. The fact that there were also a large number of failures in these family trees – nature's failed experiments – is seldom mentioned. Examples comprise the steady improvement of through-the-air transmission from Marconi to Wi-Fi and curious early attempts such as semaphores and tin cans connected by taut strings are a part of collective nostalgia; Ptolemy's epicycles and the luminiferous aether, when they are recalled, only remind us that our understanding of nature has not always been at the present level, but they are part of a narrative of success, not of failure. [1]

## 5. Implications for Communication

Our unpacking of the seeming paradox of progress despite challenges of scientific practice has implications for communication about science to society at large. There are incentives in journalism to emphasize failures rather than success. "Scientist makes modest discovery" is unfortunately about as newsworthy as "Dog bites man," whereas "Scientist makes error" appears as the punch line and headline of many reports and news articles. The effect is evident in many books and articles that have titles such as "Scientific results are not reproducible," and "How Sloppy Science Creates Worthless Cures, Crushes Hope, Wastes Billions." The problems that such publications highlight, and that scientists themselves are investigating and resolving, are not unexpected: Science is a human endeavor and resources are limited; news stories sell better if they are exaggerated. Further, the criticisms have value – they inspire improvement in the practice of science, as we presently see taking place. But when many narratives highlight problems without also discussing success and progress, it makes it easy for certain classes of readers to devalue science, and the result can be destructive (see the article in this issue by Kathleen Hall Jamieson [28] for a related view). Science does err, but that is an integral part of a process that produces valid and valuable result at the end, just as in a Darwinian biological world. The public should not be allowed to lose sight of the fact that scientific advances outweigh problems by orders of magnitude—they are responsible for substantial improvements in essentially all aspects of modern society, including health, business, industry, communication, agriculture, entertainment, and science itself. Scientific communication is an enormous field in its own right, and the many issues cannot be discussed here. We refer the reader to the perspective in this special issue by Kathleen Hall Jamieson for an interesting and more detailed exposition.

## Acknowledgments

This work as partially by the National Institutes of Health under awards P01AG039347 and U01CA198934. This work uses Web of Science data by Clarivate Analytics provided by the IU Network Science Institute.

## Footnotes

[*] The path to discovery is of course not a matter of random chance. As Pasteur famously said in 1854, "Chance favors only the prepared mind" [16]. Scientists' good judgments are the key to moving reports from 'promising' to 'valid and important' and eventually to applications.

[†] Interestingly, cold fusion sessions in some form and name have been a regular part of the American Physical Society meetings ever since.

‡ Even the invalid reports that are pursued in additional research occasionally prove useful when further research reveals the reasons for the invalidity. One example might be the false detection of gravitational waves [23], with equipment that did not have the requisite sensitivity. But the subsequent investigations showed what sensitivity would have been needed and that knowledge contributed to the subsequent search that culminated in success this year.

§ "Men become attached to certain particular sciences and speculations, either because they fancy themselves the authors and inventors thereof, or because they have bestowed the greatest pains upon them and become most habituated to them." Francis Bacon in 1620 [25].

# Recent use of the .05 criterion does not always fulfill the original intent of guarding against false assertions. In 1926, R. A. Fisher insisted upon reproducibility when he wrote [26] that he preferred "to set a low standard of significance at the 5 per cent. point, and ignore entirely all results which fail to reach this level. A scientific fact should be regarded as experimentally established only if a properly designed experiment rarely fails to give this level of significance."

| It might be tempting to describe the processes of science as Darwinian, but there is a difference. Where Nature proceeds only from generation to generation, science can retrieve – or rediscover – ideas from the past, effectively jumping from one ridge to another, as when the availability of new tools and the challenges of new problems can make a dry and unheralded theory of the past a vital tool for the present: the selection of science is not limited to the offspring of the previous generation. One example is Bayesian inference, which appeared in 1764 in a simple case and remained dormant until the 20th century when it became a flexible tool of both theoretical and applied power in the new data and computation rich environment.